\documentclass[aip,pop,reprint,groupedaddress]{revtex4-1}
\usepackage{amssymb,amsmath}
\usepackage{graphicx}

\graphicspath{ {./figures/} }
\begin{document}

\title{Modeling of Reduced Effective Secondary Electron Emission Yield from a Velvet
Surface}
\author{Charles Swanson and Igor D. Kaganovich \\
\textit{Princeton Plasma Physics Laboratory, Princeton University,
Princeton, New Jersey 08543}}
\date{October 3, 2016}
\maketitle

\section{Abstract}

Complex structures on a material surface can significantly reduce total
secondary electron emission from that surface. A velvet
is a surface that consists of an array of vertically standing
whiskers. The reduction occurs due to the capture of low-energy, true
secondary electrons emitted at the bottom of the structure and on the sides
of the velvet whiskers. We performed numerical simulations and developed an
approximate analytical model that calculates the net secondary electron
emission yield from a velvet surface as a function of the velvet whisker
length and packing density, and the angle of incidence of primary electrons.
We found that to suppress secondary electrons, the following condition on dimensionless
parameters must be met: $(\pi/2)DA \tan \theta \gg 1$, where $\theta$ is the
angle of incidence of the primary electron from the normal,
$D$ is the fraction of surface area taken up by the velvet whisker bases, and $A$
is the aspect ratio, $A \equiv h/r$, the ratio of height to radius of the
velvet whiskers.
We find that velvets available today can reduce the
secondary electron yield by $90\%$ from the value of a flat surface.
The values of optimal velvet whisker packing density that maximally
suppresses secondary electron emission yield are determined as a function of
velvet aspect ratio and electron angle of incidence.

PACS: 52.59.Bi, 52.59.Fn, 52.77.-j

\section{Introduction}

Secondary electron emission (SEE) from dielectric and metal surfaces under
bombardment of incident electron flux is important for many applications
where incident electron energy can reach tens or hundreds of electron volts.
Under these conditions secondary electron emission yield can exceed unity
and therefore strongly modify wall charging or cause multiplication of secondary
electron populations. The multipactor effect causes accumulation of SEE
population in RF amplifiers and limits the maximum electric field in these
devices \cite{multipactor}. Clouds of secondary electrons have been also
found to affect particle beam transport in accelerators. As a result,
researchers at SLAC and CERN have studied effective ways to suppress Secondary
Elecron Yield (SEY, $\gamma$)
for example by cutting grooves into the accelerator walls \cite{stupakov,
wang, pivi, suetsugu}. SEE processes are also known to affect Hall thruster
operation due to contribution to so-called near-wall conductivity or due to
reducing wall potential and increasing plasma energy losses \cite%
{raitses2011}. Wall conditions can also affect instabilities in plasmas and
electron energy distribution functions \cite{sydorenko2009, smolyakov}%
. Therefore, researchers investigate the possibility of using complex
surface structures to minimize SEY
for electric propulsion devices \cite{raitses2006, raitses2013}.

The surface geometry of a material can affect its SEY just as much as its
chemical composition. Ruzic \textit{et al.} are among those who
experimentally found that surface treatments can affect SEY \cite{ruzic}.
Alguilera \textit{et al.} studied experimentally and theoretically
velvet-covered surfaces for use in RF amplifiers to mitigate the effect of
SEY \cite{aguilera}. Cimino \textit{et al.} installed copper foam to produce
dramatic SEY reductions \cite{cimino}. Ye \textit{et al.} investigated
theoretically SEY reduction in walls with micro-pores, finding a strong
dependence on geometry and predicting as much as 45\% \ suppression of SEY
\cite{ye}. Note that there is a significant difference in micro-pores
configuration as opposed to velvet; in micro-pores configuration electrons
cannot penetrate arbitrary far along perpendicular distances into the pore
array, unlike in velvet. As an important consequence, we show that velvets can
give much higher reduction in SEY as compared to micro-pores configuration.

In this paper we study reduction in SEY by velvet surfaces both analytically
and numerically. A velvet surface is a flat substrate onto which long,
vertical whiskers are grown. The reduction of SEY comes from the fact that
low-energy true secondary electrons produced deep inside the velvet have a
large probability of hitting a whisker and getting absorbed by the surface
before exiting the velvet, therefore not contributing to net SEY from the
surface.

Baglin \textit{et al.} experimentally characterized the secondary electron
emission of dendritic copper, which has features very much like the whiskers
of velvet \cite{baglin}. They found reductions in the SEY of $>65\%$. While
we examine the effect of geometry for graphitic material, our results do
produce a similar reduction for the same velvet parameters as in Ref.\cite%
{baglin}.

More recently, Huerta and Wirz have performed Monte Carlo modeling to characterize
the SEY from copper velvet and fuzz surfaces \cite{Huerta}. They, like we do, find
reductions of SEY which have strong dependencies on the dimensionless
packing densities and aspect ratios, though they do not explore this
dependence analytically and do not cover the dynamic range of aspect ratios
that we do.

We have developed an analytical model of SEE of a velvet surface and
determined an analytical expression for the SEY, $\gamma $, of such a
surface as a function of the velvet parameters. We also simulated the SEE
process numerically and benchmarked the analytical model against the
simulation results. Based on the analytical model, we calculated approximate
values of the optimal packing density and aspect ratio of the whiskers to achieve minimum
SEY as a function of a primary angle of incidence.

Carbon velvet surfaces as available today have characteristic diameters
of a few microns, and characteristic lengths of a few
millimeters \cite{raitses2006}. The following analysis assumes that secondary
electron emission takes place only at a material surface, with no volume effects.
Because of this, our analysis holds for whisker radius $r$ larger than
the scale of primary electron penetration in carbon, tens of nanometers,

\begin{equation}
r\gg 10nm.
\end{equation}

Our analysis assumes the whiskers are grown onto a flat substrate, and that
electrons come from far away. Because of this, both the inter-whisker spacing, $s$, and the
whisker length $h$ should be much smaller than the characteristic scale length of the device, $L$:

\begin{equation}
s \ll L,
\end{equation}
\begin{equation}
h \ll L.
\end{equation}

Furthermore, our analysis assumes that velvet fibers are perfectly normally oriented. 
In actuality, velvet fibers are observed to lie at angles to the normal, and even 
to curve and change their angles, to ``flop" over near their tops \cite{raitses2006}. This effect 
is not considered by our model. 

\section{Description of Numerical Simulations of SEE Process in Velvet
Surfaces}

\label{sec:numerical}

\begin{figure}[tbp]
\centering\includegraphics[scale=0.3]{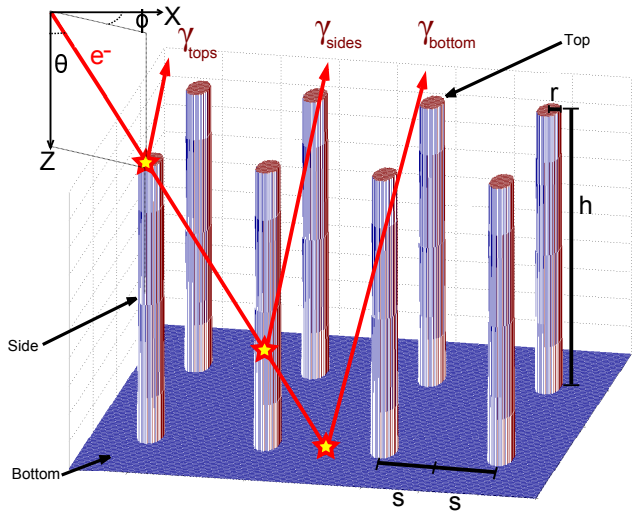}
\caption{Schematics of the velvet surface: the whisker geometrical
quantities radius, $r$, height, $h$, and spacing, $2s$. Also shown are
electron velocity polar angle $\protect\theta $, and velocity azimuthal
angle, $\protect\phi $. Numerical calculations include three contributions
to secondary electron emission from velvet: electrons emitted by the side,
top, and bottom surface of the whiskers. }
\label{figpops}
\end{figure}

In principle, it is possible to simulate electron propagation in the vacuum
and inside the material, see e.g. Ref. \cite{insepov}. However, the full
simulation is beyond necessity for our problem. Instead, we assume the SEY
of a flat surface to be known and only propagate electrons in vacuum. We
also assume that plasma does not penetrate into the whisker region because the Debye
radius is large compared with the distance between whiskers. The opposite limit
when sheath forms near two walls/surfaces and strongly affect combined SEY
of two surfaces is studied in Refs.\cite{Wang2014,Campanell2015}. When
an electron impacts the surface, SEE is produced according to known SEY of that
surface and incident angle. For a velvet surface, we have to take into
account contributions from the secondary electrons emitted by the whisker side,
top, and bottom surfaces. The electron velocity distribution function (EVDF)
is described by velocity, $\upsilon $, spherical angles, $\theta $, $\phi $,
and electron position $x,y,z$ as shown in Fig. \ref{figpops}. Geometrical
quantities of the whiskers are the whisker radius, $r$, whisker height $h$,
and spacing between whiskers, $2s$. We introduce the notation of aspect
ratio,

\begin{equation*}
A=h/r
\end{equation*}
and packing density,
\begin{equation*}
D=\frac{\pi r^{2}}{(2s)^{2}},
\end{equation*}
which is the proportion of surface area of the bottom taken up by the base
of the whiskers, see Fig. \ref{figpops}. In the numerical simulations, we
studied a regular lattice of whiskers and therefore used only one segment around
one whisker with a periodic boundary condition; particles exiting the simulation
domain were re-introduced into the opposing side with identical velocity, as if
the whiskers are arranged on a regular square grid.

\begin{figure}[tbp]
\centering\includegraphics[scale=0.5]{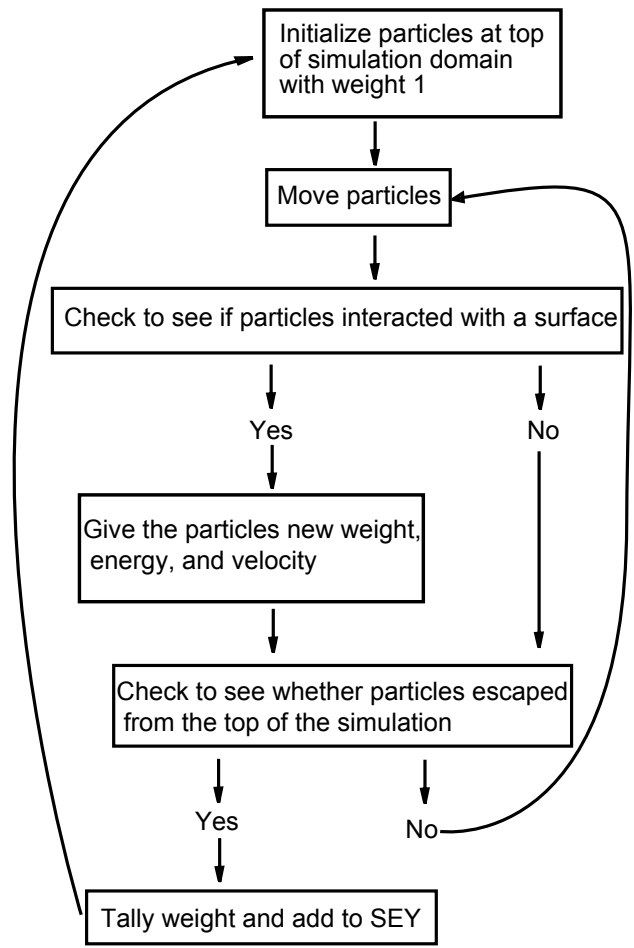}
\caption{Flowchart of the Monte Carlo simulation algorithm. }
\label{figflow}
\end{figure}

We numerically simulated the emission of secondary electrons by using the Monte Carlo method, initializing
many particles and allowing them to follow ballistic, straight-line trajectories
until they interact with surface geometry. A flowchart of the algorithm is in Fig. \ref{figflow}.
In the results presented here, we
used $10^5$ particles. Each particle object keeps track of seven quantities:
its three spatial positions $x,y,z$, its energy, $E$, and velocity angles $\theta,\phi$,
and its ``weight," meaning how many particles it stands for. All weights start
at a value of 1. Weights are changed upon interaction with a surface.

An alternate approach would be to start with fewer particles and have them stand for
a fixed number of particles, all with weight 1. When SEY occurs, in this approach one
would instantiate more particles until one tallies $10^5$. Starting with $10^5$ and instead
changing the weights upon SEY produces identical counting statistics, with error associated with
counting statistics being $N^{-1/2}=0.3\%$.

The surface geometry and initial distribution of incident particles are the simulation's main
input parameters. The velvet surface geometry is represented as cylindrical
whiskers implemented as an isosurface of a function of space, as collisions
with an isosurface are trivially detectable by a particle object which
stores its spatial location. We used an isosurface function
for which $F_{iso}=0$ defines a flat-topped cylinder with height $h$ and radius $r$
with a flat floor at $z=0$, with $F_{iso}<0$
 being inside the cylinder and $F_{iso}>0$ being outside the surface.

At every time step, we checked to see whether particles had passed
into the surface. If they had, their local normal angle was determined relative
to the gradient of $F_{iso}$, and the SEY of their energy and local normal angle
was computed. Their weight was multiplied by the SEY.

Emitted particles were given a new velocity angle. Secondary electrons were emitted with probability
linearly weighted by the cosine of the normal angle \cite{Bronstein}. Thus

\begin{equation}
P(\Omega) d\Omega=2 \cos\vartheta d\cos\vartheta \frac{d\phi}{2\pi}
\label{eq:costheta}
\end{equation}
 where
$\vartheta$ is computed relative to the local normal, $\vec{\nabla} F_{iso}$.
Specifically in the code, $\cos\vartheta=R^{1/2}$, where $R$ is a uniform
random variable from 0 to 1. The azimuthal angle in the local normal frame
was uniformly distributed from 0 to $2 \pi$.

It is interesting to note the differences between flux and velocity distribution function,
and how they characterize the number of electrons with a certain velocity. As flux is total number
of particles that pass through a differential cross sectional area oriented along some
normal $\hat{n}$, flux is

\begin{equation}
\Gamma=\int d^3v \vec{v} \cdot \hat{n} f(\vec{v})
\end{equation}

for distribution function $f(\vec{v})$. That is, flux counts particles \textit{passing through} a surface while distribution
function counts particles \textit{within} a volume. Thus, though probability and flux are
weighted by $\cos\vartheta$, this is the condition that the distribution function $f(\vec{v})$ is isotropic in angle.

The SEY of the incident electron is computed using several different \textit{a priori},
empirical, and semi-empirical expressions. We used one of the latter, that of Scholtz,
\cite{scholtz}
\begin{equation}
\gamma \left( E_{p},\theta \right) =\gamma _{max}(\theta ) \times \exp \left[ -\left( \frac{\ln [E_{p}/E_{max}(\theta )]}{%
\sqrt{2}\sigma }\right) ^{2}\right] .  \label{gamma}
\end{equation}%
where the parameters $E_{max},\gamma_{max},\sigma$ are free parameters of the Scholtz model.
Angular dependence is taken from Vaughan \cite%
{vaughan},
\begin{equation*}
\gamma _{max}(\theta )=\gamma _{max_{0}}\left( 1+\frac{k_{s}\theta ^{2}}{2\pi
}\right)
\end{equation*}
\begin{equation*}
E_{max}(\theta )=E_{max_{0}}\left( 1+\frac{k_{s}\theta ^{2}}{\pi
}\right) .
\end{equation*}

The specific constants in the Scholtz model were taken from the graphite
experimental data given in Patino \textit{et al.}\cite{patino}, $\gamma
_{max_{0}}=1.2,E_{max_{0}}=325eV,\sigma =1.6,$ $k_{s}=1$. The
semi-empirical model of Scholtz was chosen because it agrees well with
Patino \textit{et al.} experimental data for graphite.

Emitted electrons have three energy-groups. True secondary electrons were given
a low few eV temperature. Elastically scattered electrons were given the same
energy as the primary electrons. Rediffused electrons were given energy uniformly distributed
between zero and the primary electron energy. True secondary
electrons, elastically scattered electrons, and inelastically scattered
(rediffused) electrons were simulated with the energy-dependent probabilities
of emission reported in Ref. \cite{patino}. In accordance with that experiment,
 we used following empirical formula
for fraction of elastically scattered electrons, $f_{el}$:

\begin{equation}
\begin{split}
f_{el}(E_{p})&=\exp \{ 1.59+3.75\ln (E_{p})-1.37\left[ \ln (E_{p})\right]
^{2}\\+0.12\left[ \ln (E_{p})\right] ^{3}\} .
\end{split}
\label{eq:fractions}
\end{equation}%
for $E_{p}=6-390eV$, and $f_{el}(E_{p})=100\%$ for $E_{p}<6eV$, for $%
E_{p}>390eV$, $f_{el}(E_{p})=2\%$. The fraction of inelastic electrons was
assumed to be equal to $7\%$, where permitting by $f_{el}<93\%$. The values of
SEY calculated using this formula will prove to be sensitive to the fraction
of elastically scattered electrons, as these electrons are not absorbed by
surfaces, and can still contribute their full number to the SEY.

The remaining, true secondary electrons were given a Maxwellian EVDF with temperature $%
T_{true}=5.4eV$ \cite{patino}.

Simulations presented here were performed for electron incident energy of 200eV. The sensitivity
of the resulting SEY to this primary energy is not large; simulations performed at
400eV had the effect of increasing the tertiary electrons created by elastically
scattered secondary electrons, but these only account for $2\%$ of secondary electrons
at this energy.

\begin{figure}[tbp]
\centering\includegraphics[scale=0.6]{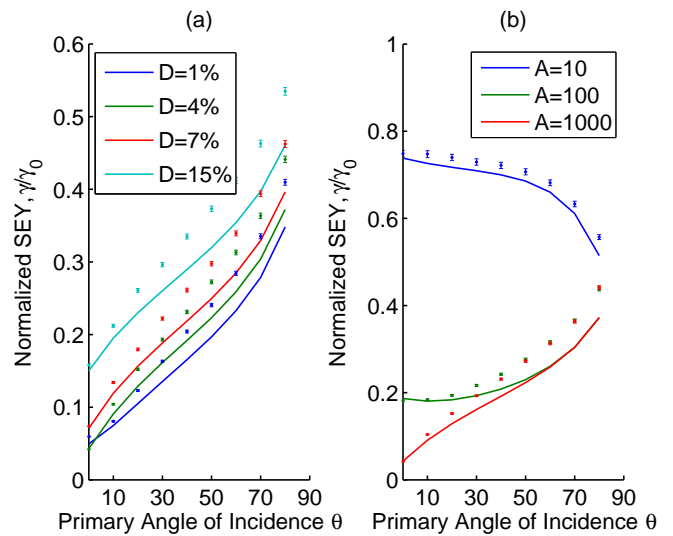}
\caption{SEY reduction from the case of normal incidence on a flat surface.
SEY reduction is given as a function of incident angle, $\protect\theta ,$ for
different values of whisker aspect ratio $A$, and packing density $D$.
Figure (a) shows SEY for 4 different $D$ values and the same $A=1000$.
Figure (b) shows SEY for 3 different values of $A$ and the same $D=4\%$.
Solid lines show the result of an analytic approximation. Points with error bars
are the result of these Monte Carlo simulations. The
simulation results are compared to the analytic approximation given in
section \protect\ref{sec:analytic}, Eq.\protect\ref{eq:final}.}
\label{figagreet}
\end{figure}

We found that adding velvet to a surface can significantly decrease the net
SEY of velvet surface as much as 90\% from the case of normal incidence on a flat surface.
 The reduction in SEY depends strongly
on the velvet parameters: packing density, $D,$ and aspect ratio, $A$. To
achieve 90\% reduction in SEY, $A=100-1000,$ and $D=4\%$ are required as
evident in Fig. \ref{figagreet}. Such velvets with aspect ratio $A=1000$
and packing density $D=4\%$ can be currently grown in the laboratory \cite%
{raitses2006}. The net SEY of velvet is a strong function of the incident
angle $\theta .$ However, for all angles the net SEY is always below 50\% of
the SEY of the flat surface without velvet for high aspect ratio $A$, see
Fig. \ref{figagreet}.

To understand the dependencies of the SEY on the velvet parameters shown in
Fig. \ref{figagreet}, an analytical model is developed and discussed in
the next section. It resolves the apparent contradiction of why the trend of
SEY with $\theta$ reverses between the values of $A=10$ and $A=100$. For the first
of these cases, the majority of the SEY is contributed by the bottom substrate. For
the second, the majority of the SEY is contributed by the sides of the cylinders.
This is described in the following analysis by the value of the dimensionless parameter
$u \tan \theta$, which crosses unity between these two cases. The significance of this
difference is explored fully in Section \ref{sec:analysis}.

\section{Analytic Expression for the Net SEY of the Velvet Surface}

\label{sec:analytic}

\begin{figure}[tbp]
\centering\includegraphics[scale=0.6]{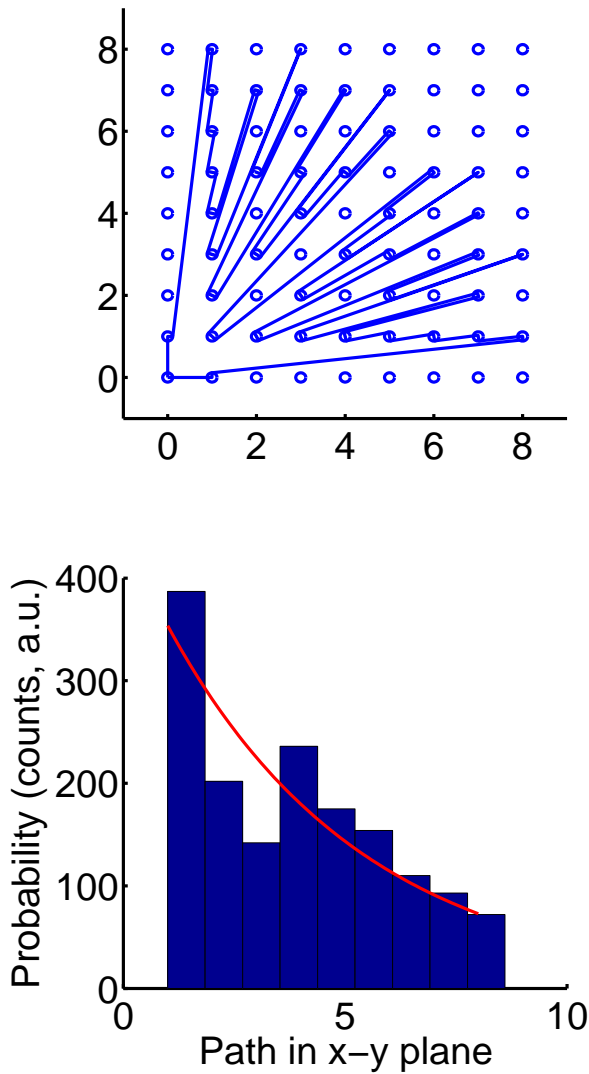}
\caption{Top: Schematics showing the top view of the velvet with regular
pattern of whiskers for $D=4\%$. Blue lines show the maximum extent of trajectories emitted at
1500 values of $\phi$, evenly spaced between 0 and $\pi/2$, originating at point $0,0$. Bottom: Probability histogram of mean-free-path for $D=4\%$ for the
rectangular case. 1500 trajectories,
evenly spaced in $\phi$, were calculated. The red curve is $\exp(-l_{\perp }/\protect\lambda %
_{\perp })$, corresponding to random configuration of whisker positions.}
\label{figrect}
\end{figure}

In the simulations we assumed that whiskers form a rectangular lattice. This
assumption helps computational tractability by introducing periodic pattern
and reducing simulation volume to one periodicity element. For the analytic
calculation it is much easier to assume instead that whiskers are positioned
along the surface randomly. We show that substituting regular lattice
whiskers instead with a random distribution of whiskers having the same
average packing density does not introduce significant change in SEY. The
probability for a secondary electron to intersect a whisker is similar for
regular and random pattern of whiskers. Qualitative agreement between
mean-free-paths given by the rectangular lattice and random lattice
assumptions is shown in Fig. \ref{figrect}. In the random position case,
this probability of not being intersecting with a whisker can be described
by exponential dependence with the distance in the $x-y$ plane
perpendicular to the whiskers' axis, $l_{\perp }$. This probability is given by

\begin{equation}
P_{free}(l_{\perp })=e^{-l_{\perp }/\lambda _{\perp }}.
\end{equation}

The mean-free-path is $\lambda_\perp$

\begin{equation}
\lambda _{\perp }=\frac{1}{2rn}.
\end{equation}

where $n=1/(2s)^2$ is the surface density of whiskers. This mean-free-path
is determined from analogy to classical hard-sphere scattering in 3D. The
cross section length that a cylinder presents is twice its radius, and the
``scattering density'' is simply the surface area density of whiskers, $n$.

From Fig. \ref{figrect} it is evident that for a rectangular lattice of
whiskers and a random configuration with the same packing density, the
mean-free-path is similar.

Our analytic model takes only one generation of true secondary electrons
into consideration. We neglect sequential secondary electrons caused by the first generation
of secondary electrons. Neglecting their contribution is the largest source of error
of the analytic, but their inclusion would add complexity to the formulae
derived. The sequential secondary electrons can be added to the treatment in the future.
The error associated with this truncation can be as high as $10\%$, as per Equation \ref{eq:fractions}

The secondary electron emission can occur on one of three surfaces: on the top of the whisker, on
the side surface, and on the bottom surface, as shown in Fig. \ref{figpops}%
. Hence the net $\gamma _{eff}$ consists of three contributions:

\begin{equation}
\gamma _{eff}=\gamma _{top}+\gamma _{bottom}+\gamma _{sides}.
\label{eqgammas}
\end{equation}

The top contribution is simply proportional to the ratio of surface of the
top whiskers to the rest of the surface. Consider a plane just below the
whisker tops. The electrons have a uniform probability of hitting any area
of that plane. Those electrons which hit the cylinder tops, whose area is $D$
of the total, will cause SEE from the tops. Those electrons which hit
elsewhere on the notional plane will penetrate into the whisker layer and
hit either the sides or the bottom surface. Therefore,

\begin{equation}
\gamma _{top}=\gamma (\theta )D.
\end{equation}

The sides and bottom contributions require calculation of probability to hit
a side surface of the whisker. As discussed above, the probability of
intersection of a side surface of the whisker can be assumed to be constant
per unit distance traveled parallel to the surface (or perpendicular to the
axis of the whisker), which is $l_{\perp }=z\tan \theta $, where $z$ is the
distance traveled along the z-axis in the velvet layer and $\theta $ is
again the angle between the z-axis and electron velocity. Thus the
probability for one electron to \textit{not intersect} a side surface of the whisker is given by

\begin{equation}
P_{free}(z,\theta)=e^{-2rnz\tan\theta}.  \label{eq:pfree}
\end{equation}

This is the cumulative probability to not have hit by depth $z$. The differential
probability to \textit{intersect} a whisker side at $z$ is given by $%
-\partial P_{free}(z,\theta )/\partial z$. The product of this and the
probability to not hit the tops $(1-D)$ is the probability that one
electron, coming in from above the whisker tops at angle $\theta$, will hit
a whisker at a height $z$

\begin{equation}
P_{hitside}(z)=-(1-D)\frac{\partial }{\partial z}P_{free}(z,\theta ).
\end{equation}

\begin{figure}[tbp]
\centering\includegraphics[scale=0.4]{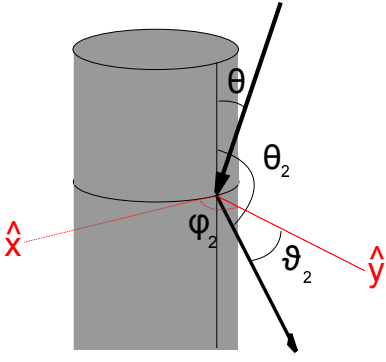}
\caption{Angle $\protect\theta $ is the polar angle of the primary electron.
Angle $\protect\theta _{2}$ is the polar angle of the secondary electron.
Angle $\protect\vartheta _{2}$ is the local normal angle of the
secondary electron. $\phi$ is the primary azimuthal angle. $\phi_2$ is the
secondary azimuthal angle. }
\label{fig:angles}
\end{figure}

To simulate how many secondaries emitted on a side surface can reach back to
free space, we need to calculate probability of emission as a function of
the velocity direction. In order to describe emitted electron velocity
direction we use the coordinate system that is placed at the whisker side as
shown in Fig. \ref{fig:angles}. In polar angles around z-axis, we
introduce angle, $\theta _{2}$ between z-axis and emitted electron velocity
and azimuthal angle, $\phi $. We also use polar angle $\vartheta
_{2}$ between local normal and emitted electron
velocity and in this same frame, a local normal azimuthal angle $\phi_2$.
 If we assume that the normal to the whisker surface coincides
with the y-axis, the $\vartheta_2$ angle between between y-axis and
emitted electron velocity is given in terms of polar angles $\theta _{2}$
and $\phi $ according to

\begin{equation}
\cos \vartheta _{2}=\sin \theta _{2}\sin \phi .  \label{costheta}
\end{equation}%
The probability distribution emitted over solid angle, $\Omega _{2}(\theta
_{2},\phi_2 )$ is

\begin{equation}
P(\Omega_2)=2\cos \vartheta _{2}\Theta (\cos \vartheta
_{2}).
\end{equation}

Here $\Theta $ is the Heviside function and we accounted for the fact that
emission flux is proportional to cosine of angle between of emitted electron
velocity direction and normal to the surface, Eq.(\ref{eq:costheta}).
Averaging over azimuthal angle $\phi_2 $ we obtain probability of electron
emission with angle $\theta _{2}$

\begin{equation}
P(\theta _{2})\equiv \int_{0}^{2\pi }\frac{d\phi }{2\pi }\sin \theta
_{2}P(\theta _{2},\phi_2 )=\frac{2}{\pi }\sin ^{2}\theta _{2} .
\end{equation}

\begin{figure}[tbp]
\centering\includegraphics[scale=0.4]{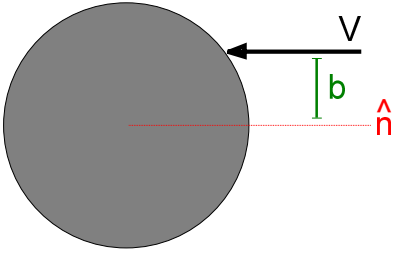}
\caption{Definition of the impact parameter, $b$. Cylinder is seen from tops,
view aligned along $z$ axis.}
\label{fig:b_int}
\end{figure}

Therefore the SEY of the sides of the velvet whiskers is

\begin{equation}
\begin{split}
\gamma _{side}(\theta )=\frac{2}{\pi }(1-D)\left\langle \gamma(\theta) \right\rangle_b  \int_{0}^{\pi /2}d\theta _{2}\sin
^{2}\theta _{2} \\ \times \int_{0}^{h}dzP_{free}(z,\theta _{2})\frac{\partial }{%
\partial z}P_{free}(z,\theta ) .
\end{split}
\label{eq:gamside}
\end{equation}

$\left\langle \gamma(\theta) \right\rangle_b$ is necessary to compute because the
local polar angle $\vartheta$ depends not only on the absolute polar angle $\theta$
but also the impact parameter at which the electron strikes (see Fig. \ref{fig:b_int}); that is an electron
which hits a fiber dead-on will have $\vartheta=\pi/2-\theta$, but an electron
which hits at a glancing angle will have $\vartheta=\pi/2$. Averaging $\gamma(\vartheta)$ over this
impact parameter $b$ gives

\begin{equation}
\left\langle \gamma(\theta) \right\rangle_b = \int_0^1 db \gamma(\cos^{-1}(\sin\theta \sqrt{1-b^2})) .
\end{equation}

Using Scholtz's and Vaughan's expressions for SEY, this averaged SEY, $\left\langle \gamma(\theta) \right\rangle_b$, was never
more than about 30\% higher than the flat value, $\gamma(\theta)$.
Substituting equation \ref{eq:pfree} into equation \ref{eq:gamside} and
introducing notation $t=\tan \theta _{2}$, equation \ref{eq:gamside} becomes

\begin{equation}
\begin{split}
\gamma _{side}(\theta )=\frac{2}{\pi }(1-D) \left\langle \gamma(\theta) \right\rangle_b
 \tan \theta \\ \times \int_{0}^{\infty }dt\frac{%
t^{2}}{(1+t^{2})^{2}}\frac{1-e^{-2rnh(t+\tan \theta )}}{t+\tan \theta }
\end{split}
\label{eq:gamsidefinal}
\end{equation}

Similarly for the contribution from the bottom surface,

\begin{equation}
\begin{split}
\gamma _{bottom}(\theta )=2(1-D)\gamma (\theta )\int_{0}^{\pi /2}d\theta_{2} \\
\times \sin \theta _{2}\cos \theta _{2}P_{free}(h,\theta _{2})P_{free}(h,\theta
) ,
\end{split}
\end{equation}

or in the same notation as equation \ref{eq:gamsidefinal},

\begin{equation}
\gamma _{bottom}(\theta )=2(1-D)\gamma (\theta )\int_{0}^{\infty }dt\frac{%
te^{-2rnh(t+\tan \theta )}}{(1+t^{2})^{2}} .  \label{eq:for_gamma_bottom}
\end{equation}

\emph{It is apparent from equations \ref{eq:gamsidefinal} and \ref{eq:for_gamma_bottom} that the dimensionless parameter }%
\begin{equation}
u=2rnh=\frac{2rh}{\left( 2s\right) ^{2}}=\frac{2}{\pi }DA  \label{eq:for_u}
\end{equation}%
\emph{is the relevant parameter to characterize a SEY from the velvet
surface.} The total SEY can be written

\begin{equation}
\gamma _{eff}(\theta )=\gamma (\theta )\left[ D+(1-D)f(u,\theta)\right] ,
\label{eq:final}
\end{equation}

where

\begin{equation}
\begin{split}
f(u,\theta )=2\int_{0}^{\infty }dt\frac{te^{-u(t+\tan \theta )}}{%
(1+t^{2})^{2}} \\+\frac{\left\langle \gamma(\theta) \right\rangle_b }{\gamma (\theta )}\tan \theta \frac{2}{\pi }%
\int_{0}^{\infty }dt\frac{t^{2}}{(1+t^{2})^{2}}\frac{1-e^{-u(t+\tan \theta )}%
}{t+\tan \theta}  .
\end{split} \label{eq:for_f}
\end{equation}

From Fig. \ref{figagreet} it is evident that the predictions of the
analytical model of velvet SEY agrees well with the numerical simulation
results. The differences are due to approximations: First, only
one generation of electrons is considered analytically. Second, the geometry simulated
is a rectangular lattice rather than the continuous distribution of scattering centers
assumed by the analytical model. The depicted simulation error derives from counting statistics,
with $\delta\gamma\propto\sqrt{N}, N=10^5$. Therefore stochastic counting error $<1\%$.

\section{Dependence of the Net SEY on Whiskers Properties }

\label{sec:analysis}

Having the analytic expression for the net secondary electron emission yield
(SEY, $\gamma _{eff}$) given by equation \ref{eq:final} allows for analysis
of optimum whisker properties for reduction of $\gamma _{eff}$. Firstly,
analysis shows that $f(u,\theta )$ is a monotonically decreasing function of
the dimensionless parameter $u$, and therefore is a monotonically decreasing
function of whisker height, see equation \ref{eq:for_u} . This is expected;
because as whisker height increases, electrons instead of hitting the bottom
surface would penetrate deeper into the whisker region further from the top
of whisker surface. Any secondary electrons produced by these electrons will
have a longer distance to traverse and large probability to hit whiskers
again and therefore net SEY is reduced.

\begin{figure}[tbp]
\centering\includegraphics[scale=0.8]{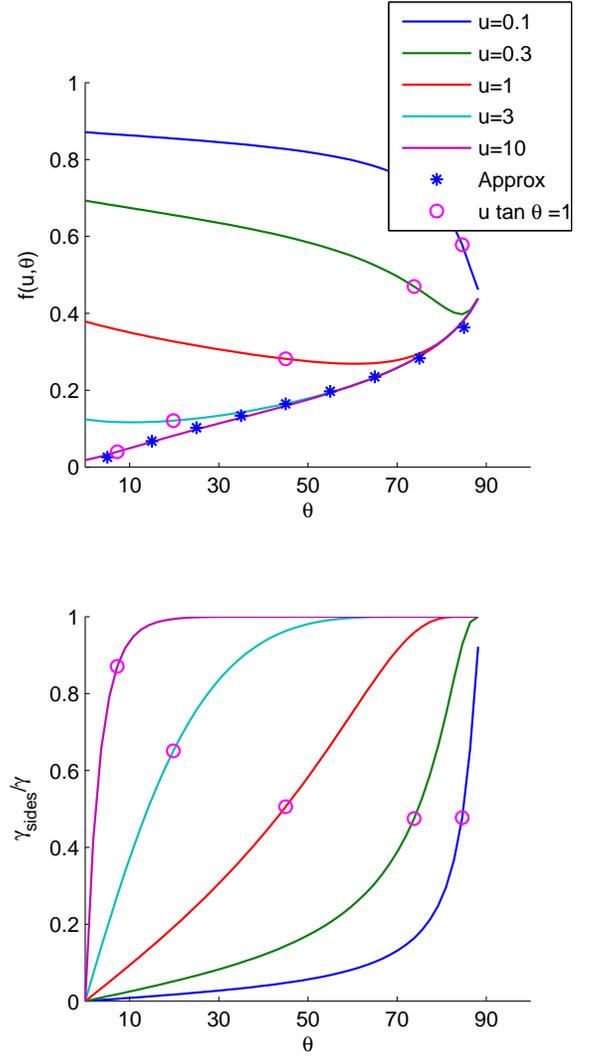}
\caption{Top: $f(u,\protect\theta )$ vs $\protect\theta $ for several $u$
(curves) that determines the net SEY in equation $\protect\gamma _{eff}=\protect%
\gamma _{flat}\{D+(1-D)f(u,\protect\theta )\}$, including the approximation
 given by equation \protect\ref{eq:approx} (blue symbols). Bottom: Relative contribution to
 the SEY of the sides of the whiskers. Pointed out in both are the points
 at which the quantity $u\tan\theta$ crosses unity.}
\label{figfut}
\end{figure}

The relative contributions from bottom and sides of whiskers is determined
by value of parameter $u\tan \theta $, as evident by comparing terms in
equation \ref{eq:for_f}. In the limit of high $u\tan \theta $
\begin{equation*}
u\tan \theta \gg 1
\end{equation*}%
the contribution of the SEY from the bottom surface into the net SEY is
negligible, because electrons hit a whisker near the tops with higher
probability. In the opposite limit
\begin{equation*}
u\tan \theta \ll 1
\end{equation*}%
the SEY from the bottom surface is significant, because electrons are
more likely to hit the bottom surface at these conditions and these
secondary electrons are more likely to escape. Moreover in this limit
contribution from the bottom surface is reduced for a more shallow angle, $%
\theta $; whereas contribution from side surfaces is increased for a more
shallow angle, $\theta .$ This is demonstrated in Fig. \ref{figfut}.

In Fig. \ref{figagreet}, the only case in which $%
u <1$ corresponds to $A=10,D=4\%$, in which $u\approx 0.25$. As
explained above the main contribution to the SEY in this case is from the bottom
surface and the net SEY is a decreasing function of the incident angle.

In the opposite limit $u\tan \theta >>1$
, the functional form of the SEY becomes

\begin{equation}
\begin{split}
\lim_{u\tan \theta \rightarrow \infty }\gamma _{eff}=\gamma (\theta )D+\frac{%
2}{\pi }\left\langle \gamma \left(\theta \right)
\right\rangle_b (1-D)\\ \times \int_{0}^{\infty }dt\frac{t^{2}\tan \theta }{%
(1+t^{2})^{2}(\tan \theta +t)} ,
\end{split}  \label{eq:gamma_limit}
\end{equation}%
which is the contribution from whisker tops and sides only (no contribution from
bottom). As evident from equation \ref{eq:gamma_limit} increasing whisker
length ($h$) above $u\tan \theta =2rnh\tan \theta >1$ will not affect the
SEY.

For the expression given by equation \ref{eq:gamma_limit} we have developed an
approximate expression in the form

\begin{equation}
\begin{split}
\lim_{u\tan \theta \rightarrow \infty }\gamma _{eff}\approx \gamma (\theta
)D+\frac{1}{2}\left\langle \gamma \left(\theta \right)
\right\rangle_b (1-D)\\ \times \left[ 1-\frac{1}{(1.39 \tan \theta +1)^{0.45}}\right] ,
\end{split} \label{eq:approx}
\end{equation}

with average deviation of 0.5\% from the exact result. This function is
depicted in Fig. \ref{figfut} (blue symbols).

\subsubsection{\protect\bigskip Optimization of velvet parameters for SEY
reduction}

\begin{figure}[tbp]
\centering\includegraphics[scale=0.3]{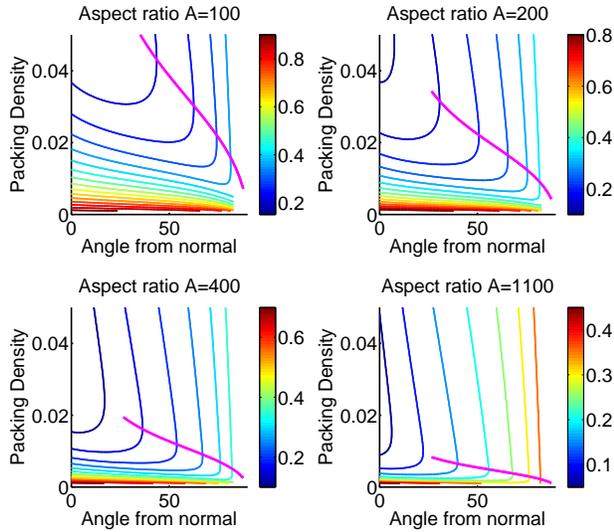}
\caption{Reduction of  Secondary Electron Yield (ratio of the net SEY to that
of a flat surface) as function of whiskers packing density and incident
angle. Magenta line shows optimal value of packing density $D(\protect\theta %
)$ corresponding to the minimum net SEY.}
\label{figcontour}
\end{figure}

In this section we investigate the velvet parameters that give SEY the most
reduction. Figure \ref{figcontour} shows SEY as a function of packing
density and incidence angle for given aspect ratio of whiskers.

From Fig. \ref{figcontour} it is evident that there is an optimal packing
density that yields the minimum net SEY for a set aspect ratio $A$. Indeed at large values of $D\rightarrow
1$ the contribution of SEY from the whisker tops dominates the net SEY  the net
SEY increases with the packing density. At small values of $D\rightarrow 0$
and large values of $A$ so that $\frac{2}{\pi }DA\tan \theta \gg 1$ the
contribution of sides dominates and
\begin{equation}
\lim_{\substack{ \frac{2}{\pi }DA\tan \theta \gg 1 \\ D\rightarrow 0}}\gamma
_{eff}\rightarrow \frac{1}{2}\left\langle \gamma \left(\theta \right)
\right\rangle_b \left[ 1-\frac{1}{(1.39\tan \theta +1)^{0.45 }%
}\right]   \label{eq:gamma_sides}
\end{equation}%
and for very small values of $D\rightarrow 0$ such that $\frac{2}{\pi }%
DA\tan \theta \ll 1$ contribution of the bottom of whiskers dominates and
taking appropriate limit in equation \ref{eq:final} gives
\begin{equation}
\lim_{\substack{ \frac{2}{\pi }DA\tan \theta \ll 1 \\ D\rightarrow 0}}\gamma
_{eff}\rightarrow \gamma (\theta  ) .
\end{equation}%
That is at very low values of  $D$ there are simply not enough whiskers for
any significant number of electrons to intersect them when traversing the
velvet.

Therefore the optimum value of whisker parameters corresponding to minimum
SEY is approximately given by the condition

\begin{equation}
\frac{2}{\pi }DA\tan \theta \sim 1 .
\end{equation}

Investigation of Fig. \ref{figcontour} allowed us to derive the location of
the optimal $D$ to minimize net SEY, $\gamma _{eff}$ for given values of $%
A,\theta $. This optimal $D_{optimal}$ is approximately given by

\begin{equation}
D_{optimal}(\theta )\approx 0.97\frac{\ln (A)}{A(\tan \theta )^{0.47}}-\frac{%
0.26}{A} . \label{eq:Dopt}
\end{equation}

Equation \ref{eq:Dopt} gives agreement with 12\% average error for the
position of the optimal D as shown in Fig. \ref{figcontour}.

\section{Conclusions}

We have investigated numerically and analytically the effect of velvet surfaces
on secondary electron emission (SEE) and concluded that the net  secondary
electron yield (SEY) can be reduced dramatically by application of velvet to
the surface. Geometrical quantities of the whiskers are the whisker radius, $%
r$, whisker height, $h$, and spacing between whiskers, $2s$. The beneficial
velvet configuration for the net SEY reduction consists of high aspect ratio
long whiskers,

\begin{equation}
A\equiv h/r ,
\end{equation}

 rarely placed on the surface with low packing
density

\begin{equation}
D\equiv \pi r^{2}/(2s)^{2}\rightarrow 0 ,
\end{equation}

 such that
\begin{equation*}
\frac{2}{\pi }DA\tan \theta \gg 1 .
\end{equation*}%
In this case incident electrons do not reach the bottom of the velvet and
large fraction of the secondary electrons emitted from the side do not exit
the velvet, because they are intersected by whiskers again. The approximate
net SEY is given in this case by
\begin{equation}
\lim_{\substack{ \frac{2}{\pi }DA\tan \theta \gg 1 \\ D\rightarrow 0}}\gamma
_{eff}\rightarrow \frac{1}{2}\left\langle \gamma \left(\theta \right)
\right\rangle_b \left[ 1-\frac{1}{(1.39\tan \theta +1)^{0.45}}%
\right]   \label{eq:gamma_limit1}
\end{equation}

where

\begin{equation}
\left\langle \gamma(\theta) \right\rangle_b = \int_0^1 db \gamma(\cos^{-1}(\sin\theta \sqrt{1-b^2})) .
\end{equation}

From equation \ref{eq:gamma_limit1}  it is evident that it is possible to
decrease SEY by more than 50\% for shallow incidence ($\tan \theta \sim 1$)
and more than 90\% for normal incidence ($\theta \ll 0.7$) compared to the case of
normal incidence on a flat surface.

The optimal packing density for reducing SEY depends on the angle of
incidence of the primary electrons and is approximately given by
\begin{equation}
D_{optimal}(\theta )\approx 0.97\frac{\ln (A)}{A(\tan \theta )^{0.47}}-\frac{%
0.26}{A} .
\label{eq:Dopt1}
\end{equation}

Equation \ref{eq:Dopt1} gives agreement with 12\% average error for the
position of the optimal D as shown in Fig. \ref{figcontour}.

In summary, using plausible values for parameters of lab-grown velvets ($%
A>100$), we find that velvet surfaces are a promising candidate for reducing
SEY by more than 50\% for shallow incidence ($\tan \theta \sim 1$) and more
than 90\% for normal incidence ($\theta \ll 0.7$) compared to the case of normal incidence on a flat surface.
The closer to the normal
is the angle of incidence, the more a long, non-dense velvet can suppress
the SEY. Interestingly, if it is known that the incident electron flux has a
\textit{narrow} distribution of incident angles, one could use a velvet with
whiskers oriented not normally but along that direction; such velvet is the
most efficient way to minimize SEY.

\subsection{Acknowledgment}

The authors would like to thank Alexander Khrabrov for checking the math and ensuring
consistency of the distribution function and flux formulations.
This work was conducted under a subcontract with UCLA with support of AFOSR under grant
FA9550-11-1-0282.

\end{document}